\documentclass[conference]{IEEEtran}
\IEEEoverridecommandlockouts
\usepackage{cite}
\usepackage{amsmath,amssymb,amsfonts,amsthm}
\usepackage{graphicx}
\graphicspath{{./figures/}}
\usepackage{subfigure}
\usepackage{textcomp}
\usepackage{mathtools}
\usepackage{multirow}
\DeclarePairedDelimiter\ceil{\lceil}{\rceil}

\usepackage[utf8]{inputenc}
\usepackage[english]{babel}
\usepackage{url}

\usepackage{xcolor}
\usepackage{algorithm,algcompatible,amsmath}
\algnewcommand\INPUT{\item[\textbf{Input:}]}
\algnewcommand\OUTPUT{\item[\textbf{Output:}]}

\def\BibTeX{{\rm B\kern-.05em{\sc i\kern-.025em b}\kern-.08em
    T\kern-.1667em\lower.7ex\hbox{E}\kern-.125emX}}
\begin{document}

\title{Scheduling Deep Learning Jobs in Multi-Tenant GPU Clusters via Wise Resource Sharing}

\author{
\IEEEauthorblockN{Yizhou Luo\IEEEauthorrefmark{1}, Qiang Wang\IEEEauthorrefmark{2}, Shaohuai Shi\IEEEauthorrefmark{2}, Jiaxin Lai\IEEEauthorrefmark{1}, Shuhan Qi\IEEEauthorrefmark{2}, Jiajia Zhang\IEEEauthorrefmark{2}, Xuan Wang\IEEEauthorrefmark{2}\thanks{Corresponding authors: Qiang Wang, Shaohuai Shi}}

\IEEEauthorblockA{Harbin Institute of Technology (Shenzhen)}
\IEEEauthorblockA{Guangdong Provincial Key Laboratory of Novel Security Intelligence Technologies}
\IEEEauthorblockA{\IEEEauthorrefmark{1}\{23S151149,200110515\}@stu.hit.edu.cn, \IEEEauthorrefmark{2}\{qiang.wang,shaohuais,shuhanqi,zhangjiajia,wangxuan\}@hit.edu.cn}

}
\maketitle

\begin{abstract}
Deep learning (DL) has demonstrated significant success across diverse fields, leading to the construction of dedicated GPU accelerators within GPU clusters for high-quality training services. Efficient scheduler designs for such clusters are vital to reduce operational costs and enhance resource utilization. While recent schedulers have shown impressive performance in optimizing DL job performance and cluster utilization through periodic reallocation or selection of GPU resources, they also encounter challenges such as preemption and migration overhead, along with potential DL accuracy degradation. Nonetheless, few explore the potential benefits of GPU sharing to improve resource utilization and reduce job queuing times. 

Motivated by these insights, we present a job scheduling model allowing multiple jobs to share the same set of GPUs without altering job training settings. We introduce SJF-BSBF (shortest job first with best sharing benefit first), a straightforward yet effective heuristic scheduling algorithm. SJF-BSBF intelligently selects job pairs for GPU resource sharing and runtime settings (sub-batch size and scheduling time point) to optimize overall performance while ensuring DL convergence accuracy through gradient accumulation. In experiments with both physical DL workloads and trace-driven simulations, even as a preemption-free policy, SJF-BSBF reduces the average job completion time by 27-33\% relative to the state-of-the-art preemptive DL schedulers. Moreover, SJF-BSBF can wisely determine the optimal resource sharing settings, such as the sharing time point and sub-batch size for gradient accumulation, outperforming the aggressive GPU sharing approach (baseline SJF-FFS policy) by up to 17\% in large-scale traces.
\end{abstract}


\section{Introduction}
The popularity of Deep Neural Network (DNN) \cite{lecun2015deep} grows rapidly in both industry and academia with its significant role in various applications, such as computer vision and natural language processing. 
With more and more training data and larger model size, training deep models becomes very time-consuming. 
Distributed Deep Learning (DDL) \cite{lsddn2012} is widely adopted to speed up the training procedure, which distributes the training workload to a cluster of workers and exploit the parallel computing power to accelerate the training process. 

In the data center scenario where the hardware resources are shared by multiple users, multiple online DDL training jobs are running simultaneously, and the resource contention could lead to severe performance degradation if the training jobs are not scheduled properly \cite{jeon2019analysis}. For such an online scheduling system that concurrently handles a rising number of jobs, flexible resource allocation and efficient job scheduling are indispensable to maximize the resource utilization. There exist some traditional schedulers \cite{plan2015,network2019,tetri2016,liu2019,makespan2019} to schedule different computing tasks, but they are not specifically designed for DDL training jobs and cannot leverage the characteristics of DDL (such as iterativeness and convergence properties) for maximal training efficiency. 

Existing DL job management and scheduling systems \cite{xin2017scheduling, lowPrep2019, tony2019, optimus2018, tiresias2019, gavel_osdi2020} commonly employ preemptive and exclusive strategies to enhance system utilization and minimize job completion time. The advanced heuristic scheduler Tiresias \cite{tiresias2019} demonstrated that the shortest-remaining-service-first (SRSF) algorithm generally yields optimal results when job durations are known. However, small jobs still experience delays waiting for GPU resource release when the cluster is predominantly occupied by large jobs.
The state-of-the-art representative is Pollux \cite{aurick2021_pollux}, which dynamically (re-)assigns resources to improve cluster-wide goodput, while respecting fairness and continually optimizing each DL job to better utilize those resources. However, Pollux helps users choose the GPU resources as well as tune the training hyper-parameters, which may result in model accuracy degradation \cite{blox_2024}. 
Overall speaking, in preemptive and exclusive policies, long-term job packing can exacerbate HOL (Head-of-line) blocking issues and prolong JCT (Job Completion Time). Consequently, jobs with small training iterations and low GPU demand may face severe queuing and starvation issues, while those large ones can suffer from high migration overhead.

In several recent schedulers, including Gandiva \cite{gandiva2018}, Zico \cite{gangmuk_zico_2021}, Salus \cite{sayed2019_salus} and Lucid \cite{hu2023_lucid}, there has been a notable shift towards emphasizing resource sharing, particularly regarding GPU and network resources. This shift aims to enhance overall resource utilization while addressing queuing and starvation issues effectively.
Gandiva \cite{gandiva2018} introduced GPU time-slicing and job scheduling based on predicted DDL training job characteristics, albeit with a conservative approach limiting GPU sharing to single-GPU jobs.
Yu et al. \cite{yu2022_gadget,yu2022_rar} tackled network resource sharing in multiple all-reduce based DDL job training, reducing communication contention overhead.
Lucid \cite{hu2023_lucid} utilized an indolent packing strategy to mitigate interference effectively. 
However, their search was confined to a limited solution space due to the inability to alter the training batch size.

On the contrary, gradient accumulation has become standard feature in many deep learning training frameworks \cite{grad_accum_ccgrid2023}. The basic idea of gradient accumulation is to accumulate gradients from multiple micro-batches and only then update the model parameters. This is particularly helpful in training very large neural networks \cite{huang2019_gpipe}, where workers can only fit one small micro-batch at a given time, saving GPU memory footprint requirement. From an optimization perspective, gradient accumulation is completely equivalent to training with a larger mini-batch size, since in both cases the gradient is averaged with respect to all computed examples.

Motivated by the observations outlined above, we introduce a job scheduling model that enables multiple jobs to run concurrently on one or more GPUs. In contrast to the approach of scaling the training batch size and tuning training hyper-parameters as in \cite{aurick2021_pollux}, we investigate the potential of GPU sharing to improve the overall performance. This model is coupled with gradient accumulation to address GPU memory limitations and ensure model convergence. The contributions of this paper can be summarized as follows:
\begin{itemize}
    \item We introduce a novel DDL job scheduling model enabling multiple jobs to fully or partially share the same set of GPUs while ensuring model convergence through gradient accumulation. Unlike existing methods that increase batch size and GPU numbers to enhance performance, risking accuracy degradation, our model focuses on GPU resource sharing across jobs and mitigates GPU memory constraints through gradient accumulation, thereby potentially reducing queuing time for waiting DDL jobs.
    \item We propose SJF-BSBF (shortest job first with best sharing benefit first), a straightforward yet effective scheduling algorithm for the aforementioned problem. Initially, we derive the optimal solution for scheduling a job pair (one ongoing job and one new arrival) to decide GPU sharing feasibility and launch timing. Subsequently, we employ a greedy strategy to determine batch size and GPU allocation, minimizing interference with existing jobs and reducing queuing time.
    \item Through both physical and simulated experiments, we evaluate SJF-BSBF on different scales of job traces. Compared to recent DL schedulers such as Tirasias \cite{tiresias2019} and Pollux \cite{aurick2021_pollux}, SJF-BSBF reduces average job completion time by 27-33\%. Additionally, compared to the first-fit GPU sharing approach for new arrival jobs, SJF-BSBF avoids those sharing decisions that may degrade the overall performance, surpassing it by up to 17\%.
\end{itemize}


\section{Related Work}\label{sec:relatedwork}

Scheduling DL training jobs has garnered significant interest recently. Research in this field primarily focuses on fully utilizing computing resources and allocating them effectively to achieve optimal efficiency in multi-tenant GPU computing environments. Here we discuss two categories: preemptive and exclusive schedulers, as well as non-preemptive schedulers.

\textbf{Preemptive and Exclusive Schedulers.}
These schedulers possess the capability to interrupt or preempt a running job in order to allocate exclusive resources to another job with higher priority. This mechanism ensures that allocated resources remain inaccessible to other jobs while the current job is utilizing them, thereby fostering predictable resource usage patterns and mitigating interference between jobs.
Early works such as Optimus \cite{optimus2018} and Cynthia \cite{cynthia2019} relied on job time prediction, making simplistic assumptions about training convergence curves. Tiresias \cite{tiresias2019} addressed the severe resource starvation issue by proposing adaptive scheduling algorithms with effective job migration strategies. Other studies, such as Harmony \cite{dlblb2019} and Spear \cite{spear2019}, leveraged deep reinforcement learning to provide efficient solutions aimed at minimizing average job completion time or makespans.
Another line of research in DDL job scheduling algorithms relies on theoretical formulation and optimization, treating DDL job scheduling as constrained optimization problems. The recent state-of-the-art Pollux \cite{aurick2021_pollux} dynamically reallocates resources to enhance cluster-wide throughput while ensuring fairness and continually optimizing each DL job to maximize resource utilization.
However, these methods cannot guarantee no accuracy degradation for all models. Moreover, they may encounter performance degradation due to migration \cite{tiresias2019} and GPU under-utilization \cite{weng2022_mlaas}.

\textbf{Non-preemptive Schedulers.}
Early non-preemptive schedulers predominantly relied on heuristic algorithms based on job characterization and hardware performance modeling. In recent studies, attention has shifted towards resource sharing, encompassing GPU and network resources, which holds significant potential for improving computing resource utilization and alleviating starvation.
Gandiva \cite{gandiva2018} introduced GPU time-slicing and job scheduling by predicting DDL training job characteristics. However, it adopted a conservative approach, limiting GPU sharing to single-GPU jobs.
Zico \cite{gangmuk_zico_2021} focused on system-wide memory consumption for concurrent training and devised a feasible memory management solution to ensure that concurrent jobs do not exceed the allocated memory budget.
Wang et al. \cite{wang2020communication} and Yu et al. \cite{yu2022_gadget,yu2022_rar} addressed network resource sharing in multiple ring-all-reduce based DDL job training, alleviating communication contention overhead. 
Lucid \cite{hu2023_lucid} employed an indolent packing strategy to mitigate interference. However, few of these approaches offer a general and flexible solution for sharing GPUs among DL jobs.


\section{Preliminaries}\label{sec:preliminaries}
\subsection{S-SGD Based Distributed Deep Learning}\label{bg:ddl}
The DNN model is trained in an iterative manner with the target of minimizing a loss function $\mathcal{L}\left(W, D\right)$, where $W$ and $D$ are respectively the model weights and the input data. For large-scale DNNs, the data-parallel synchronized SGD (S-SGD) is widely applied to train models with multiple workers (say $N$ workers, and indexed by $g$) because it has the same convergence performance as the sequential SGD. Generally the ${i^{th}}$ iteration of the training contains four steps: a)  Each worker $g$ loads a mini-batch of local data $D_{i}^g$ into the device memory. b) Each worker $g$ performs a feed forward on $D_{i}^g$ through the neural network and computes the value of the loss function $\mathcal{L}(W_i, D_{i}^g)$. c) The first order gradients w.r.t. $W_i$ are calculated by backpropagation. d) Gradients from all the workers $\nabla\mathcal{L}(W_i, D_{i}^g)$ are aggregated, averaged and then distributed, which is often tackled by the All-Reduce collective function. Then all the workers update the model as Eq. \eqref{eq:ssgd_update}. 
\begin{align}
    W_{i+1} = W_{i}-\xi \frac{1}{N}\sum_{g=1}^{N}\nabla\mathcal{L}(W_i, D_{i}^g). \label{eq:ssgd_update}
\end{align}


\subsection{All-Reduce Communication}\label{subsec:ar_comm}
The most common scenario of DDL training is using a large number of computing devices distributed among nodes in a cluster. As a result, the step d) involves extra communication overheads 
In Eq. \eqref{eq:ssgd_update}, we use $\Delta W_i = \frac{1}{N}\sum_{g=1}^{N}\nabla\mathcal{L}(W_i, D_{i}^g)$ to represent the aggregation of gradients from $N$ workers, which can be done through an all-reduce operation or through a set of parameter servers. 
For brevity, we assume that the number of nodes is power-of-two. Given the number of nodes $N$ and the message size $M$, the time cost of one All-Reduce operation without contention can be generalized as where $a$ and $b$ are two constant numbers that are not related to $M$ \cite{model_all_reduce}. The inter-node communication cost can be modelled as Eq. \eqref{eq:t_comm}. The values of $a$ and $b$ depend on the algorithms for the All-Reduce operation with different number of processes and message sizes \cite{model_all_reduce}. Without loss of generality, we do not limit the communication model to one specific algorithm. 
\begin{align}
	T_{allreduce}=a + bM. \label{eq:t_comm} 
\end{align}

\section{System Modeling and Problem Formulation}\label{sec:problemformulation}
For ease of reference, we summarize some frequently used notations throughout this paper in Table \ref{tab:notation}. 
\begin{table}[!ht]
	\centering
	\vspace{-1.2 em}
	\caption{Frequently used notations}
	\vspace{-0.8 em}
	\begin{tabular}{|c|l|} \hline
		Name & Descriptions \\ \hline \hline
		$\mathcal{S/N}$ & Set of servers/GPUs in the cluster \\
		\textsl{$S_{i}$} & the $i$-th server of the cluster \\
		\textsl{$g_{i,j}$} & the $j$-th GPU on the $i$-th server \\
		\hline
		$\mathcal{J}$ & The job set \\
		$G_j$ & \# of GPUs requested by job $j$ \\
		$J_{k}$ & the $k$-th job \\
		$\mathcal{G}(J_{k})$ & the set of GPUs used by $J_{k}$ \\
		$\mathcal{S}(J_{k})$ & the set of servers used by $J_{k}$ \\
		$a_{k}$ & the arrival time of $J_{k}$ \\
		$B_{k}$ & the mini-batch size per-GPU used by $J_{k}$ \\
		$I_{k}$ & \# of iterations that $J_{k}$ needs to run \\
		$t_{k}$ & The execution time of one iteration of $J_{k}$ \\
		$L_{k}$ & The total execution time of $J_{k}$ running solely \\
		$T_{k}$ & The completion time of job $j$ \\
		$E_{k}$ & the timestamp when $J_{k}$ is finished \\
		\hline
		$\mathcal{F}$ & The set of feasible scheduling solution \\
		${f}_{j}^{k}$ & The schedule of job $j$ in the $k$-th solution \\
		\hline
	\end{tabular}
	\label{tab:notation}
\end{table}

We consider a multi-tenant GPU cluster comprising $|\mathcal{S}|$ servers equipped with $|\mathcal{N}|$ GPUs evenly distributed. These servers are interconnected with a network switch possessing sufficient bandwidth. All GPUs within the cluster share the same specifications and theoretical peak performance. At the onset of a scheduling horizon $|\mathcal{T}|$ spanning time-slots, a set of DDL jobs $\mathcal{J}$ awaits scheduling for training over the duration of $|\mathcal{T}|$. Each job $J_k \in \mathcal{J}$ is characterized by the number of GPUs it requires, denoted as $\mathcal{G}(J_k)$, and the total number of training iterations $I_k$ requested by its users.

\subsection{DL Job Training Time Modeling}
We first model the training time of one job, which includes the GPU computation time and network communication of the all-reduce operation. 
\subsubsection{Modeling GPU Computation}
The DL model is trained using back-propagation. The computation time on GPU scales linearly with the per-GPU batch size $B$, which can be calculated as follows.
\begin{equation}
	t_{comp}(B)={\alpha}_{comp} + {\beta}_{comp} \times B. \label{eq:comp}
\end{equation}

\subsubsection{Modeling Network Communication}
The gradient aggregation overhead depends on the topology as well as the network communication algorithm. We simply define the communication part as follows.
\begin{equation}
	t_{comm}={\alpha}_{comm} + {\beta}_{comm} \times M, \label{eq:comm}
\end{equation}
where $M$ is the message size, and $\alpha_{comm},\beta_{comm}$ are the all-reduce time model parameters as described in Section \ref{subsec:ar_comm}.
\begin{figure}[!h]
	\centering
	\includegraphics[width=0.98\linewidth]{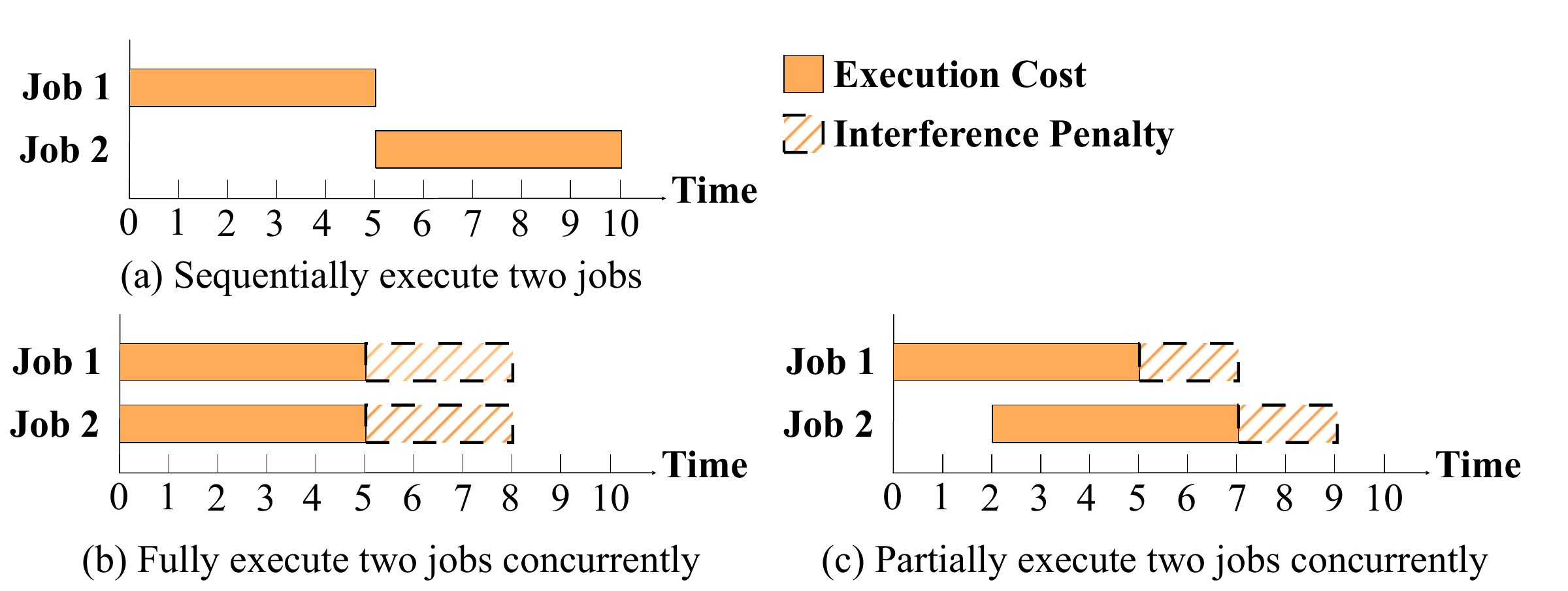}
	\caption{Three job schedules for two DL jobs.}
	\label{fig:share_interfere}
\end{figure}
\subsubsection{Sharing Performance Modeling}
Existing schedulers that facilitate GPU sharing, such as Gandiva \cite{gandiva2018}, Gavel \cite{gavel_osdi2020}, and Lucid \cite{hu2023_lucid}, often adopt conservative and limited approaches or require additional application information to generate schedules. In contrast, we apply a simple interference model to describe the overhead of GPU sharing. We illustrate three possible job schedules for two jobs sharing the same set of GPUs in Figure \ref{fig:share_interfere}. Schedule (a) sequentially executes two DL jobs. Schedules (b) and (c) involve invoking two DL jobs simultaneously or with partial overlap, resulting in varying degrees of interference penalty.
To optimize the average job completion time, one must balance the tradeoff between job queuing/waiting time (Job 2 waits for Job 1 to finish in (a)) and interference penalty (complete overlap of two jobs leads to severe penalty in (b)). In practice, the job iteration time under GPU sharing can be measured and modeled by equations \eqref{eq:comp} and \eqref{eq:comm}, as they occupy partial GPU and network resources with similar trends. To simplify the model, if a new job shares GPUs occupied by an existing job (Job $A$ and Job $B$), we adjust their job iteration time as follows.
\begin{align}
	\hat{t}_{A} &= t_A {\xi}_A,  \\
	\hat{t}_{B} &= t_B {\xi}_B,  
\end{align}
where ${\xi}_A$ and ${\xi}_B$ denote the interference ratios, reflecting the performance degradation resulting from GPU sharing. The solution to determining the optimal scheduling point under this scenario will be discussed in Section \ref{subsec:one_pair}.

\subsubsection{Modeling Gradient Accumulation}
Given that GPU memory constraints may limit the per-GPU batch size, some schedulers tackle this limitation through memory offloading \cite{swapadvisor_asplos2020} (which may introduce additional system overhead) or by adjusting batch sizes and other training hyper-parameters \cite{aurick2021_pollux} (which may compromise model accuracy). As our model incorporates GPU sharing, the memory footprint frequently imposes constraints on feasibility. Thus, we focus on gradient accumulation, which can dynamically reduce the sub-batch size while preserving the original model accuracy as per the user's requested batch size. It is also easily implemented using popular DL frameworks. It is important to note that one can utilize gradient accumulation algorithms to manage the computational aspect, thereby reducing batch size to mitigate memory consumption. We subsequently define the overall iteration time as follows.
\begin{align}
    t_{iter}^{j}=(s-1) \times t_{comp}^j(\frac{B}{s}) &+ (({t_{comp}(\frac{B}{s})})^{\delta}+{t_{comm}}^{\delta})^{1/\delta}, \label{eq:job_iter}
\end{align}
where $s$ represents the accumulation step required to attain the original batch size, and $\delta$ denotes the degree of overlap between GPU computation and all-reduce communication, as initially proposed in \cite{aurick2021_pollux}. It is important to acknowledge that $\delta$ may vary when different batch sizes are applied.

\subsection{Scheduling Modeling}
In this paper, we adopt the "gang-scheduling" discipline widely prevalent in practical large-scale GPU clusters \cite{tiresias2019,themis2020,yu2022_rar}. Under gang scheduling, all workers (i.e., GPUs) of a DDL job must be allocated simultaneously. Furthermore, once a job commences its scheduled run, all allocated GPUs must remain dedicated to the job until its completion, with no allowances for preemption or migration. (It is worth noting that frequent job preemption and migration can significantly degrade performance \cite{tiresias2019}). Upon job completion, the occupied resources are simultaneously released. Differing from conventional GPU-exclusive scheduling policies, we permit GPUs to be occupied by multiple workers from various jobs concurrently. These workers can be allocated within a single server or across multiple servers, provided there exists a network path connecting them.

Assume $y_{jg}[\tau]$ denotes that the job $j$ uses GPU $g$ in the time slot $t$. 
Job $j$ requires $G_j$ number of GPUs. 
\begin{align}
    \sum_{g \in \mathcal{G}}y_{jg}[\tau] = G_j. 
\end{align}

To ensure that one GPU at most holds $C$ jobs, we have
\begin{align}
	\sum_{j \in J[\tau]}y_{jg}[\tau] \leq C,  \forall j \in \mathcal{J}[t], \tau \in \mathcal{T}, g \in \mathcal{G}.
\end{align}
In practice, we observe that interference degradation can be severe, rarely improving performance when more than two jobs share the same set of GPUs. Therefore, we set $C=2$ in our context.

Also, since we consider gang scheduling, we have
\begin{align}
    y_{jg}[\tau]&=y_{gs}[\tau-1], \forall s \in S, j \in \mathcal{J}[\tau], a_j < \tau \leq T_j, \\
    y_{jg}[\tau]&=0, \forall g \in \mathcal{G}, j \notin \mathcal{J}[\tau], \tau \in \mathcal{T}, \\
    y_{jg}[\tau]& \in {\mathbb{Z}}^{+}, \forall g \in \mathcal{G}, j \in \mathcal{J}[\tau], \tau \in \mathcal{T}.
\end{align}

The completion time of job $j$ can be calculated as
\begin{align}
    T_j = a_j + \text{arg }\underset{\tau}{\text{min}}& {\sum_{\tau \in \mathcal{T}} \frac{1}{t_{iter}^{j}} \geq I_k}, \forall j \in \mathcal{J}[\tau], \tau \ge a_j, \\
    &{\phi}_j[t] = \frac{B_k}{t_{iter}^{j}}, \label{eq:thrpt}
\end{align}
where ${\phi}_j[\tau]$ denotes the system throughput of the job.
\begin{figure*}[!ht]
	\centering
	\subfigure[ImageNet]
	{
		\includegraphics[width=0.3\linewidth]{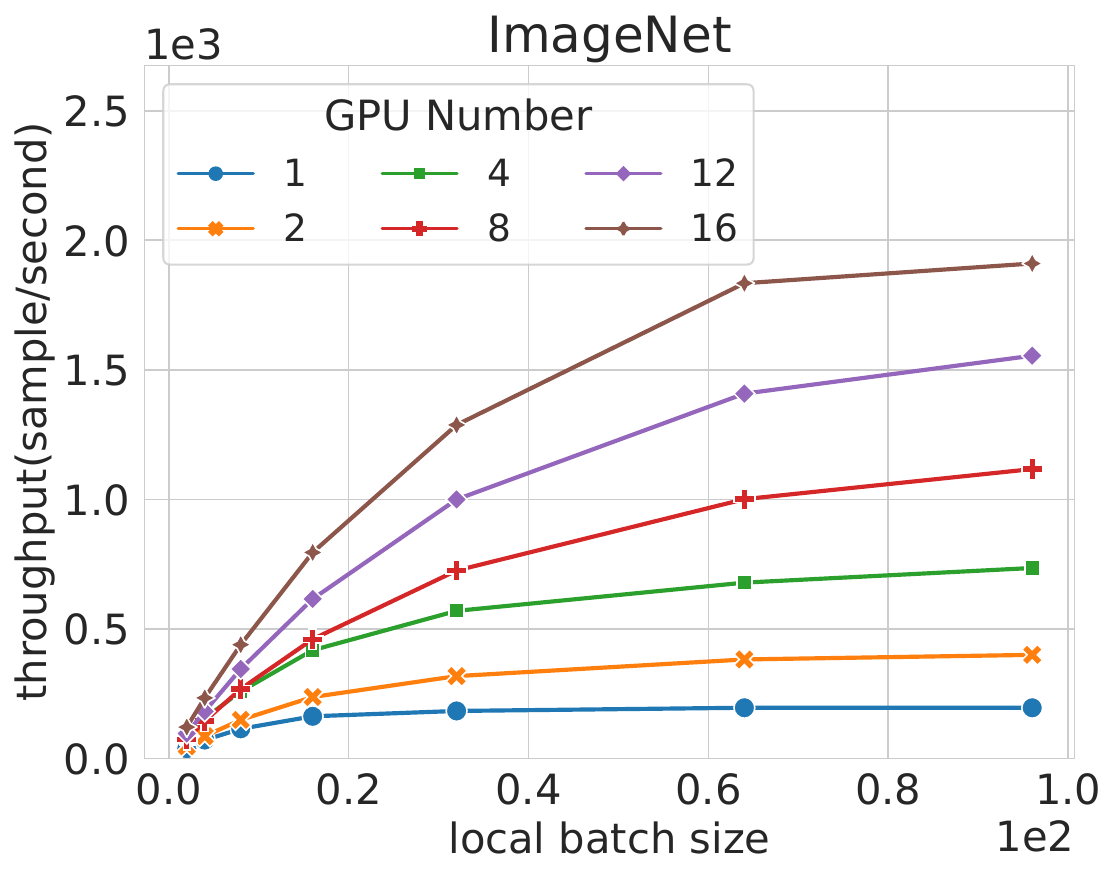}\label{fig:single_imagenet}
	}
	\subfigure[YoloV3]
	{
		\includegraphics[width=0.3\linewidth]{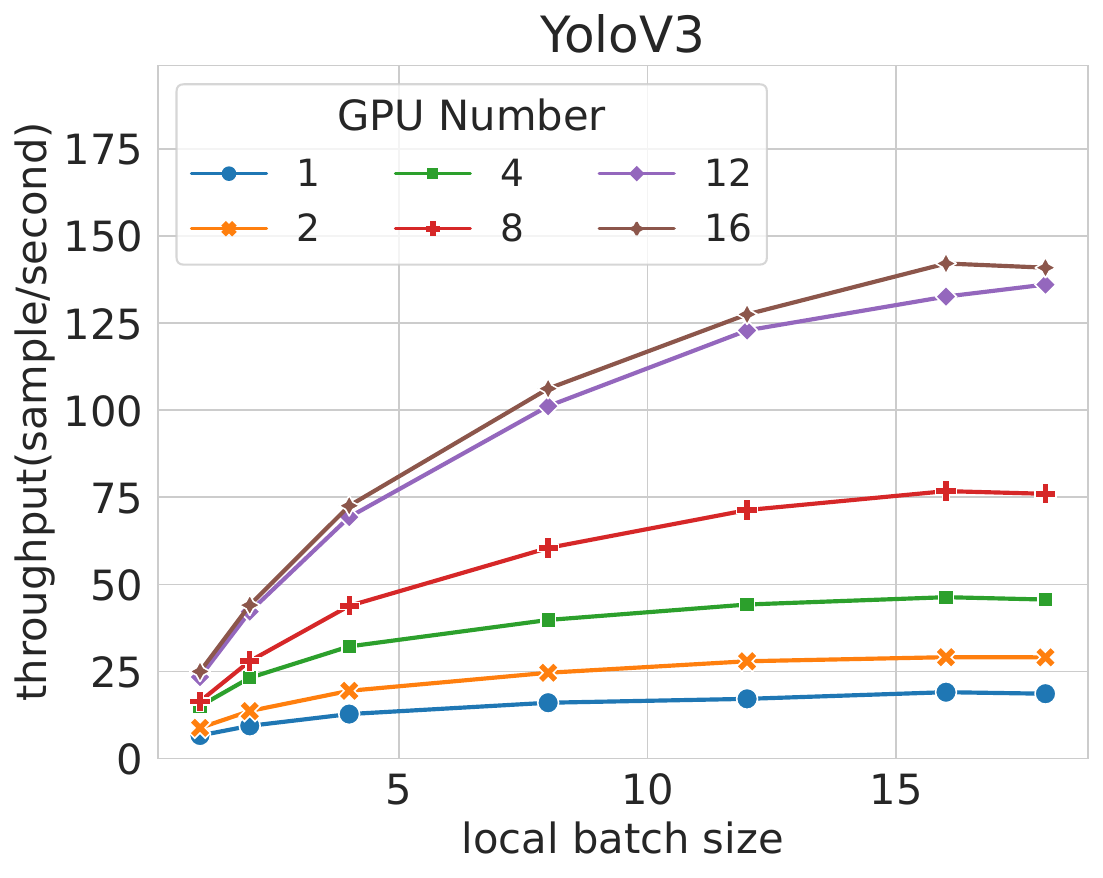}\label{fig:single_yolov3}
	}
	\subfigure[DeepSpeech2]
	{
		\includegraphics[width=0.3\linewidth]{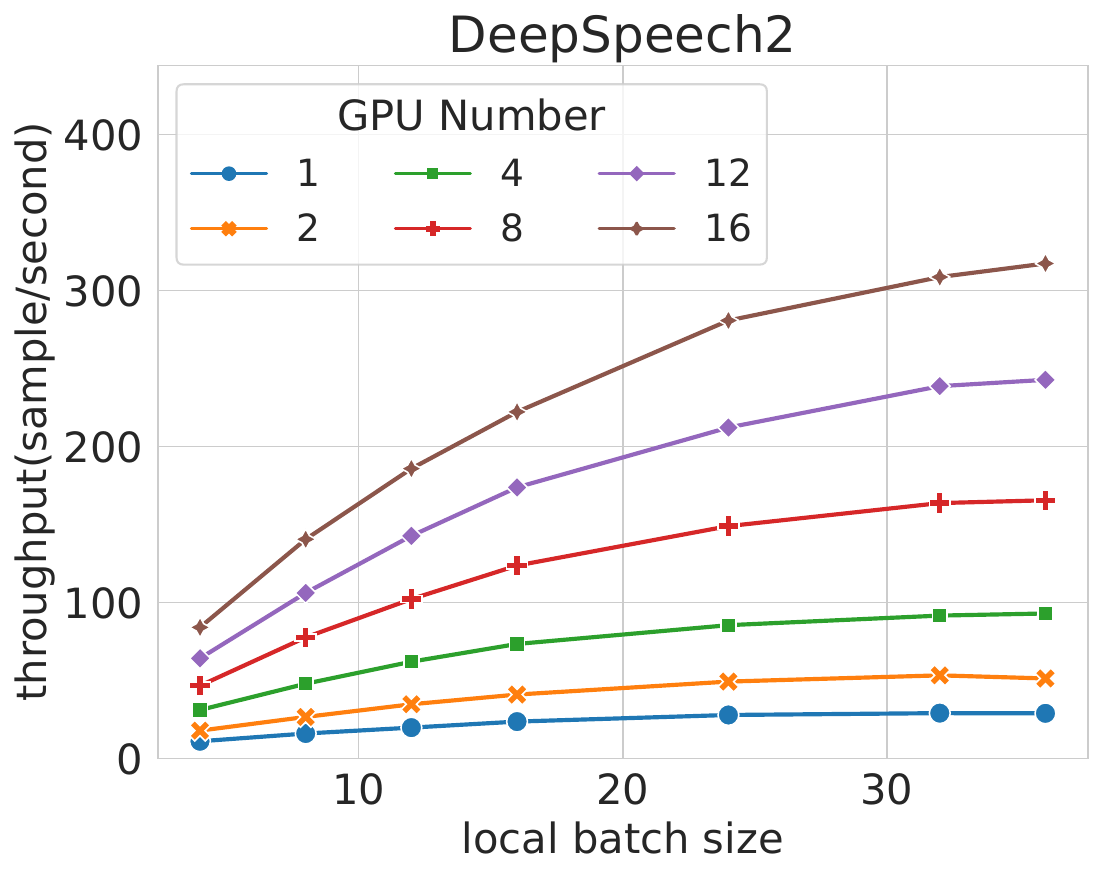}\label{fig:single_ds}
	}
	\subfigure[BERT]
	{
		\includegraphics[width=0.3\linewidth]{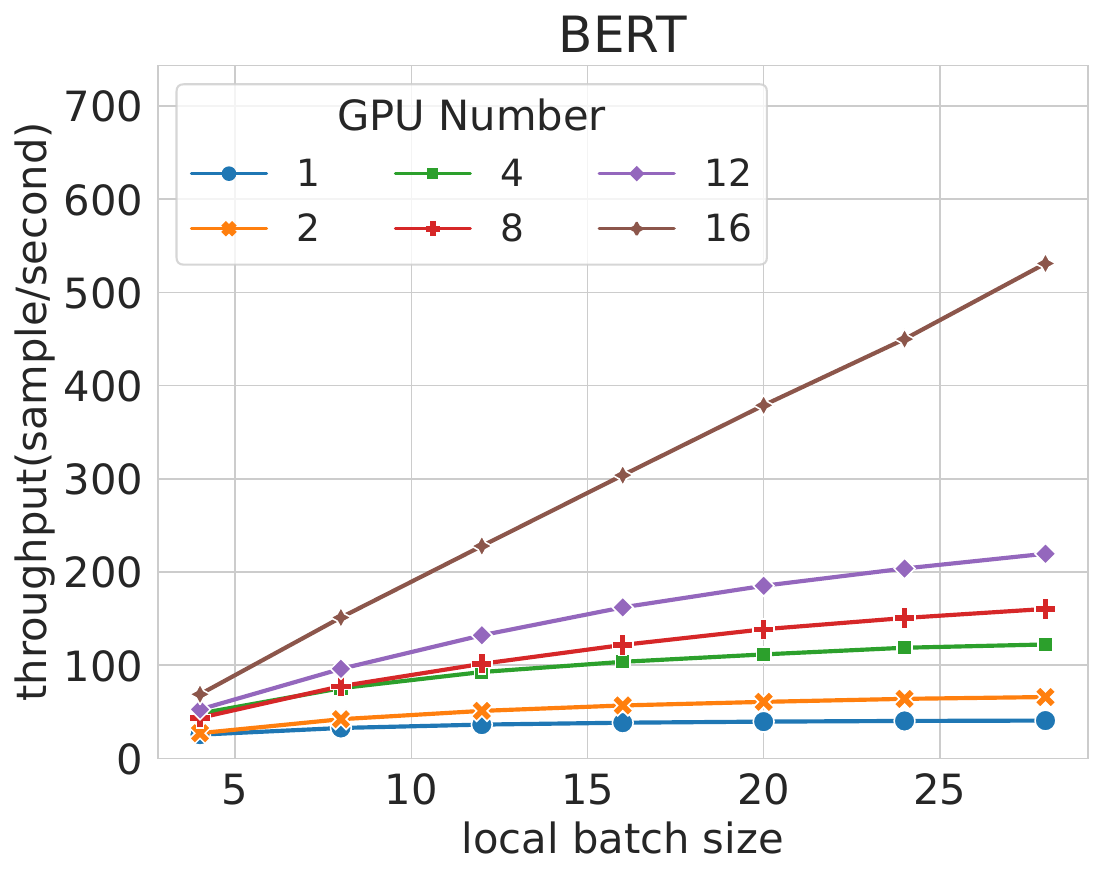}\label{fig:single_bert}
	}
	\subfigure[CIFAR10]
	{
		\includegraphics[width=0.285\linewidth]{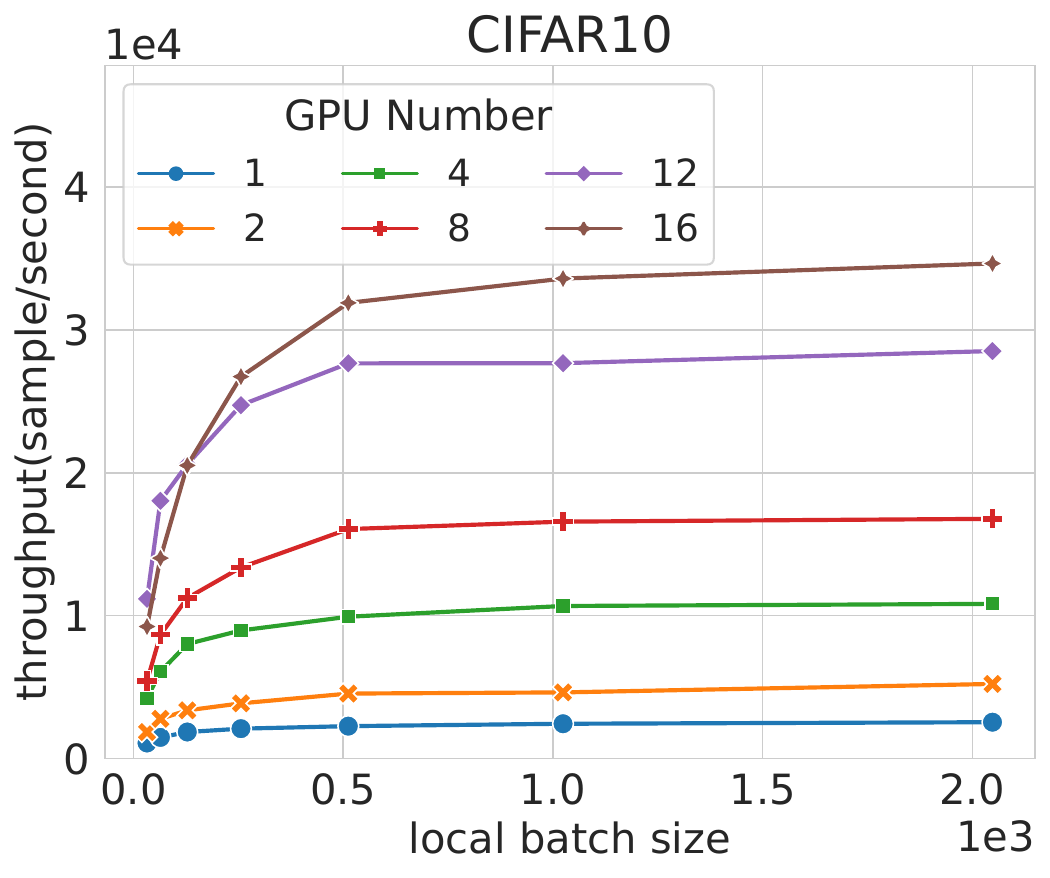}\label{fig:single_cifar10}
	}
	\subfigure[NEUMF]
	{
		\includegraphics[width=0.3\linewidth]{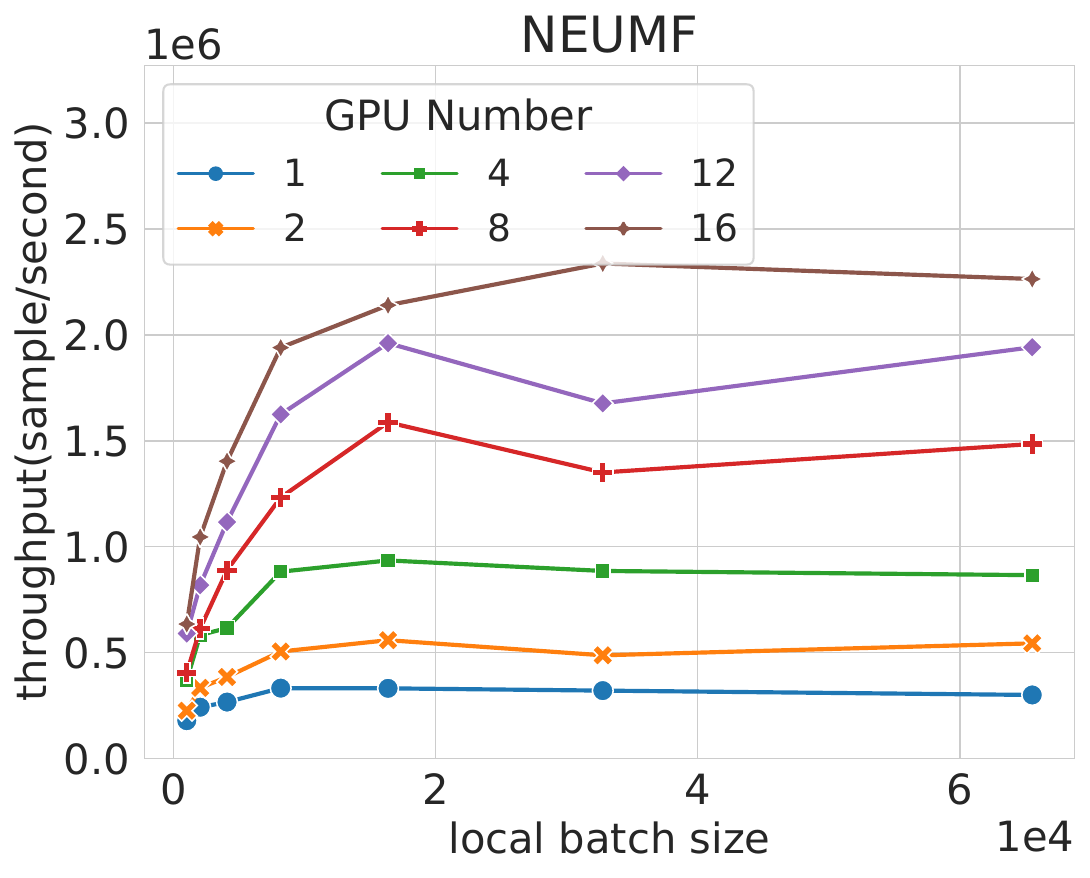}\label{fig:single_neumf}
	}
	\caption{System throughput for all DL models in our experiments, as measured using a 4-server cluster each with 4 NVIDIA 2080Ti GPU. Each sub-figure shows the values of different resource and training batch size settings for each model.}
	\label{fig:single_thrpt}
\end{figure*}
\begin{figure*}[!ht]
	\centering
	\includegraphics[width=0.92\linewidth]{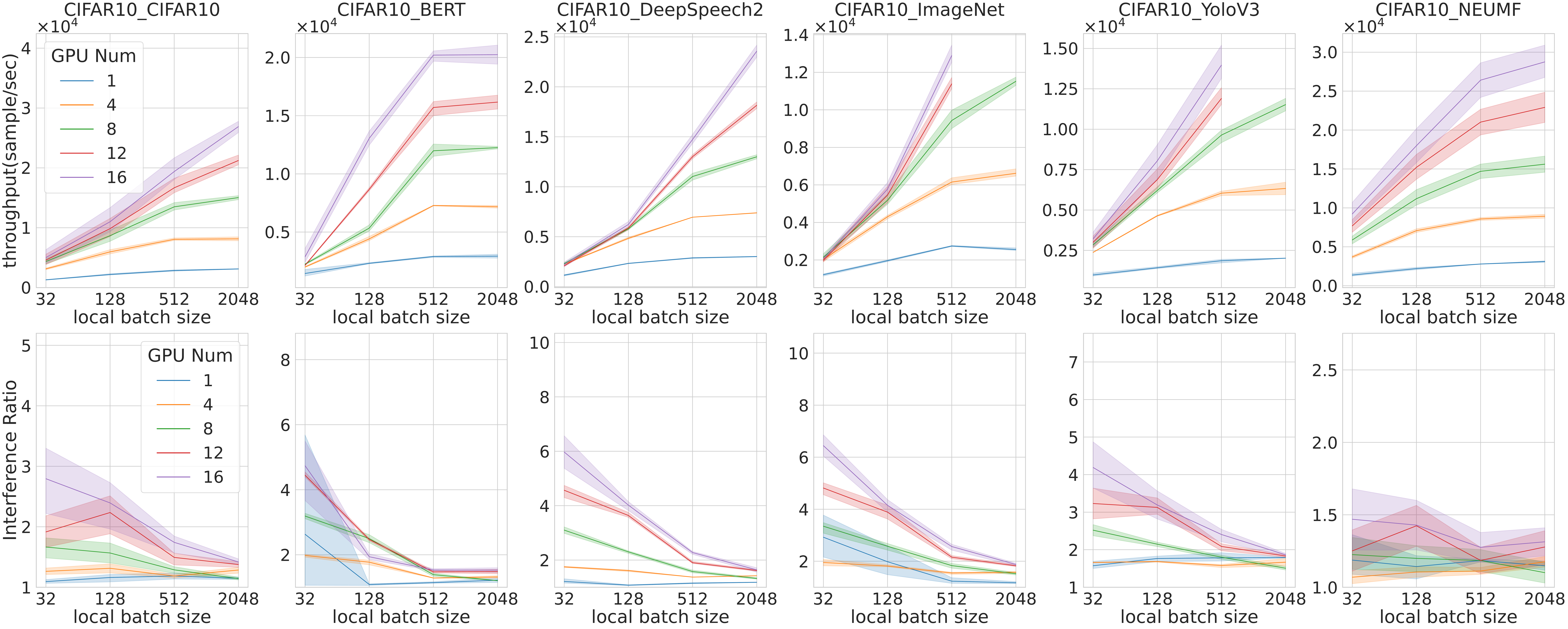}
	\caption{TOP: System throughput of difference DL models paired with CIFAR10 to share the same set of GPUs. BOTTOM: the interference ratio $\xi$ for different DL models and resource and training settings.}
	\label{fig:twin_thrpt}
\end{figure*}
In practice, it is more common to monitor and collect DL training throughput using popular DL frameworks. The throughput can be readily converted to iteration time given the training batch size. By measuring DL job throughput under both sole execution and concurrent execution with other jobs, we can fit the time model (Equation \eqref{eq:job_iter}) for both cases and naturally infer the interference ratio $\xi$. Figure \ref{fig:single_thrpt} illustrates the throughputs of all DL models in our experiments across a range of resource allocations and batch sizes. Overall, our model closely represents the observed data. We also notice that different jobs exhibit varying sensitivities to network communication and GPU workloads. For instance, BERT shows a linear increase with batch size within the experimental range for all GPU configurations, indicating that the bottleneck lies in GPU computation and is constrained by GPU memory. Additionally, YoloV3 mostly achieves peak throughput with a batch size of 16 and encounters network bottlenecks when the GPU number exceeds 12.
We also measure the system throughput of different job pairs and training configurations, as depicted in Figure \ref{fig:twin_thrpt}. We find that throughput can be fitted by Equation \eqref{eq:thrpt}, albeit with different parameters from the solely running mode. Moreover, the interference ratios of different cases exhibit a wide range of up to 6 in our experiments, emphasizing that avoiding unfavorable cases is crucial for improving overall performance.

\subsection{Problem Formulation}
In this paper, our goal is to determine the scheduling decisions $y_{jg}[t]$ to minimize the average JCT, which is commonly used to evaluate the efficiency of DL job schedulers \cite{tiresias2019,aurick2021_pollux}. This optimization problem can be formulated as follows:
\begin{align}
    \underset{y_{jg}[t],\forall j,g,t}{min}\sum_{j \in J[t]} T_j. \label{prob:jct}
\end{align}

\section{Solution}\label{sec:solution}
We note that Problem \eqref{prob:jct} presents an integer non-convex program with packing and covering constraints, which is NP-hard. Given these challenges, we opt to explore a heuristic approach that provides a provable local optimum guarantee for a job pair that shares GPUs either completely or partially.

In this section, we describe our solution to address the problem formulated in Section \ref{sec:problemformulation}. The solution comprises two parts.
Firstly, we address the simple case of two jobs: one running on the GPUs while the other awaits scheduling. It is important to note that concurrent execution on a GPU may degrade overall performance if interference is significant. Thus, we must decide whether the jobs should share GPUs and when to launch the waiting one. This gives rise to \textbf{Theorem 1}, which forms the core of our solution by providing a feasible solution when cluster resources are insufficient.
Secondly, we introduce our scheduling algorithm SJF-BSBF (shortest job first with best sharing benefit first), built upon \textbf{Theorem 1} and the shortest job first strategy. By judiciously selecting job pairs that benefit from GPU sharing, even acting in a non-preemptive manner, SJF-BSBF reduces job queuing time while avoiding scenarios where sharing may detrimentally impact overall performance.

\subsection{Scheduling One Job Pair}\label{subsec:one_pair}
We assume that all the tasks of a DL job are assigned to a fixed set of GPUs during its execution.
Before we design the scheduling algorithm, each new-arriving DL job should be placed to a certain set of intra-node or inter-node processors, which is called job placement. 

Assume that there is a new job $A$ sharing the GPUs occupied by the existing job $B$, and their execution time under concurrent execution is respectively
\begin{align}
    \hat{t}_{A} &= t_A {\xi}_A,  \\
    \hat{t}_{B} &= t_B {\xi}_B,   
\end{align}
and $\kappa$ is the inserting time. We have the following theorem.

\textbf{Theorem 1} The shortest JCT of the above job pair is achieved by either sequentially executing them ($\kappa=t_A i_A$) or simultaneously invoking them concurrently $\kappa=0$. 
\begin{proof}
Case 1: If $\hat{t}_{A}i_A \geq \hat{t}_{B}i_B$, then
\begin{align}
    T_A &= \hat{t}_{B}i_B + t_A \times (i_A - \frac{\hat{t}_{B}i_B}{\hat{t}_A}), \\
    T_B &= \hat{t}_{B}i_B. 
\end{align}
The average time is
\begin{align}
    \overline{T} &= (T_A + T_B) / 2, \\
                 &= \hat{t}_{B}i_B + \frac{t_A i_A}{2} - \frac{\hat{t_B}i_B}{2{\xi}_{A}}.
\end{align}
Case 2: If $\hat{t}_{A}i_A < \hat{t}_{B}i_B$, then
\begin{align}
    T_A &= \kappa + \hat{t}_{A} \times (i_A - \frac{\kappa}{t_A}), \\
    T_B &= \kappa + \hat{t}_{A} \times (i_A - \frac{\kappa}{t_A}) + t_B \times (i_B - \frac{\hat{t_A} \times (i_A - \frac{\kappa}{t_A})}{\hat{t_B}}).
\end{align}
The average time is
\begin{align}
    \overline{T} &= (T_A + T_B) / 2 \\
                 &= (\frac{2{\xi}_{B}+{\xi}_{A}-2{\xi}_{A}{\xi}_{B}}{2{\xi}_{B}})\kappa + (1-\frac{1}{2{\xi}_{B}})\hat{t_A}i_A + \frac{1}{2}t_B i_B.
\end{align}

If $2{\xi}_{B}+{\xi}_{A}-2{\xi}_{A}{\xi}_{B} > 0$, it is an monotonically increasing function with respect to $\kappa$. The minimum value is achieved at $\kappa=0$, which indicates that one should start the new job immediately.
Otherwise, it is an monotonically increasing function with respect to $\kappa$. The minimum value is achieved at $\kappa=t_A i_A$, which indicates that overlapping two jobs degrade the overall performance.
\end{proof}

In practice, evaluating the conditions for the best solution is the same as directly comparing the fully overlapped time and the fully non-overlapped time in terms of time cost. 

\subsection{Scheduling Multiple DL Jobs}
One critical challenge in addressing Problem \eqref{prob:jct} is the allocation of GPUs when all GPUs in the cluster are occupied by existing jobs, and a new job arrives. One needs to determines 1) which job to share resources with the new arrival, 2) deciding when to initiate the new job. Our goal is to develop a heuristic efficient online scheduling algorithm for it.
\subsubsection{\textbf{Basic Idea}} We propose an online scheduling algorithm called SJF-BSBF (\underline{s}mallest \underline{j}ob \underline{f}irst with \underline{b}est \underline{s}haring \underline{b}enefit \underline{f}irst). Algorithm \ref{algo:sjf_bsbf} describes the steps of SJF-BSBF. 
The intuition behind Algorithm \ref{algo:sjf_bsbf} has three points. 
1) As for job priority, the overall framework is based on the shortest job first (SJF) strategy, as tackled by Lines 1-2. This size-based heuristic strategies determine the job priority according to their used GPU numbers. We apply SJF since it performs well most of time for online scheduling problems \cite{tiresias2019}. 
2) As for GPU allocation, since the case of scheduling two jobs concurrently running on the same GPUs (wholely or partially) is considered in our paper, once the free GPU number is not enough to execute the job, we manage to look for those already occupied by the running jobs to schedule the new one. This is the core logic of SJF-BSBF and handled by Lines 3-19.
3) In the point of 2), for each job pair, we should also decide the batch size of the new job for gradient accumulation that not only exceeds the GPU memory size but also achieves the shortest JCT of scheduling the job pair. This corresponds to Line 11.
\begin{algorithm}
	\caption{Shortest Job First with Best Sharing Benefit First (SJF-BSBF)}
	\label{algo:sjf_bsbf}
	\begin{algorithmic}[1]
		\INPUT The pending job list $\mathcal{J}_{pending}$ to allocate GPUs at the current scheduling time point. $L_k$ is the expected remaining time of $J_k$, calculated by $t_{iter}^j \times I_k$. $\mathcal{G}_{OJ}$ denotes the GPU set occupied by at least one job. $\mathcal{J}_{share}$ denotes the candidate job set for concurrent execution.
		\OUTPUT $\mathcal{G}(J_{k})$, The GPU sets of each job $J_{k}$ in $\mathcal{J}_{pending}$.
		\STATE Sort $\mathcal{J}_{pending}$ by $L_{k}$ in ascending order.
		\FOR {$J_k$ in $\mathcal{J}_{pending}$}
		\STATE $\mathcal{G}(J_{k}) \leftarrow \emptyset$.
		\STATE $\mathcal{G}_{free} \leftarrow$ GPUs that hold no job.
		\STATE $\mathcal{G}_{OJ} \leftarrow$ GPUs that hold one job. 
		\IF {$|\mathcal{G}_{free}| \ge G_{J_k}$}
		\STATE $\mathcal{G}(J_{k}) \leftarrow $ the top-$G_{J_k}$ GPUs in $\mathcal{G}_{free}$
		\ELSE
		\IF {$|\mathcal{G}_{free}|+|\mathcal{G}_{OJ}| \ge G_{J_k}$}
		\FOR {$g$ in $\mathcal{G}_{OJ}$}
		\STATE Get $SF, b, t$ using Algorithm \ref{algo:bs_scale}.
		\STATE Add $\mathcal{J}_g$ to $\mathcal{J}_{share}$ \textbf{if} {SF is True}.
		\ENDFOR
		\STATE Sort $\mathcal{J}_{share}$ by $t$ in ascending order.
		\FOR {$J$ in $\mathcal{J}_{share}$}
		\STATE Add $\mathcal{G}_{J}$ to $\mathcal{G}_{J_k}$ until {$|\mathcal{G}_{J_k}| \ge G_{J_k}$}.
		\ENDFOR
		\ENDIF
		\ENDIF
		\ENDFOR
	\end{algorithmic}
\end{algorithm}

\subsubsection{\textbf{GPU Allocation}}
Given a job $J_k$ that needs $G_{J_k}$ GPUs, we should decide a set of GPUs to schedule it. The classical heuristic algorithms include First-Fit (FF) \cite{scheduling2011} and List-Scheduling (LS) \cite{scheduling2011}. In Algorithm \ref{algo:sjf_bsbf}, Lines 3-19 present the step of choosing the GPUs. First, if there are enough GPUs to execute $J_k$, we select the top-$k$ GPU in $\mathcal{G}_{free}$ to make them as colidated on the nodes as possible (Lines 6-7). Second, notice that we allow at most two jobs to concurrently run on the same GPUs. Once the free GPU number is smaller that the request of $J_k$, we attempt to seek those GPUs that are already occupied by one job (Lines 10-17). We scan $G_{OJ}$ and determine the best concurrent running setting of the running job and $J_k$, including the batch size of $J_k$ and whether to let them share the GPUs, using Algorithm \ref{algo:bs_scale} introduced later. We add those pairs that can benefit from the sharing strategy (Lines 10-13). Then we sort $\mathcal{J}_{share}$ by the JCT of the job pair in ascending order (Line 14). Finally, we pick up the GPUs from those candidate jobs until the total number can fulfill the request of $J_k$ (Lines 15-17). Notice that we do not pick the free GPUs at first for this case to save resources because the completion time of $J_k$ is determined by those shared GPUs. 
\begin{algorithm}
	\caption{Batch Size Scaling with Best Sharing Benefit}
	\label{algo:bs_scale}
	\begin{algorithmic}[1]
		\INPUT A given job $J_k$ to allocate GPUs. The user's requested batch size $B_k$. $t_{share}$ represents the JCT of the best sharing configuration of two jobs. 
		\OUTPUT The sharing configuration denoted by $\overline{SF}, \overline{b}, \overline{t}$. $\overline{SF}$ indicates whether to share the GPU for the new job.
		\STATE $b \leftarrow B_{J_k}$
		\STATE $\overline{b} \leftarrow B_{J_k}$
		\STATE $\overline{t} \leftarrow \text{a large number}$
		\STATE $\overline{SF} \leftarrow False$ 
        \WHILE{$\ceil*{b} \neq 1$}
        \STATE Get $SF$ and $t_{share}$ according to \textbf{Theorem 1}.
        \IF {$t_{share} \le \overline{t}$}
        \STATE $\overline{t} \leftarrow t$
        \STATE $\overline{b} \leftarrow b$
        \STATE $\overline{SF} \leftarrow SF$
        \ENDIF
        \STATE $b \leftarrow b/2$
        \STATE return $\overline{SF}, \overline{b}, \overline{t}$.
        \ENDWHILE
	\end{algorithmic}
\end{algorithm}
\subsubsection{\textbf{Batch Size Scaling}}
In Algorithm 2, given a running job $J_{run}$ and a new job $J_{k}$ ready to be scheduled, we present how to adaptively adjust the batch size for $J_{k}$ to achieve the shortest average JCT of these two jobs. Notice that we do not adjust the batch size of the running job to reduce the complexity of the scheduling system. We search the batch size in the range [1, $B_{J_k}$] with a step of power two (Lines 5 and 12). For each candidate batch size, we use \textbf{Theorem 1} to obtain the best configuration of scheduling the job pair, including the flag of whether to let them share GPUs ($SF$) and the JCT (Line 6). Then we record that configuration if better (Lines 7-11). Notice that it is possible that the new job $J_{k}$ may not be scheduled immediately and put back to the pending job pool if the final $\overline{SF}$ is $False$, indicating that running the job pair concurrently is not optimal.

\subsubsection{\textbf{Time complexity of SJF-BSBF}} 
The time consumption of SJF-BSBF is primarily attributed to searching the GPU set for a pending job to be shared when there is insufficient resource (Lines 9 to Line 18 in Algorithm \ref{algo:sjf_bsbf}). Initially, a for loop (Line 10) scans all GPUs with one job, iterating $|\mathbb{G}(OJ)|$ times. Subsequently, each iteration executes Algorithm \ref{algo:bs_scale} to determine the time point to initiate GPU sharing as well as the appropriate batch size, with a time complexity of $\theta$(${\log}2 (B{J_k}))$. Lastly, after collecting candidate jobs for sharing GPUs, sorting the list in ascending order to select those with the shortest Job Completion Time (JCT) requires $\theta(|\mathcal{J}{share}|{\log}2(|\mathcal{J}{share}|))$. Consequently, the time complexity for scheduling a job is $\theta(|\mathbb{G}(OJ)|{\log}2 (B{J_k})+|\mathcal{J}{share}|{\log}2(|\mathcal{J}{share}|))$. In our system implementation on a 16-GPU cluster, the overhead of periodically scheduling those waiting jobs is negligible, averaging below 0.02 seconds for each operation.

\section{Performance Evaluation}\label{sec:experiments}
\subsection{Experimental Setup}


\textbf{Cluster configurations:} We first conduct physical experiments on a cluster of four servers. Each server is equipped with an Intel Xeon CPU E5-2670 and four Nvidia GeForce 2080 Ti GPUs. The network is configured with Fat-Tree topology with 10 Gbps connected to a 100-Gbps switch. All experiments are performed in the environment of Ubuntu 20.04, PyTorch 1.18, CUDA 11.2 and OpenMPI 4.0. Based on the data measured on the physical environment, we then conduct simulation experiments to resemble the physical cluster configuration and test large scale of clusters and job traces.
To evaluate the performance of SJF-BSBF in a large-scale cluster (16 servers each with 4 GPUs) with long-term traces, we also implement a simulator to record job events and resource usage. 
All experiment results without explicit comments are derived from the simulation.

\textbf{Baselines:} We consider the following baselines.
\subsubsection{First-In-First-Out (FIFO)}
a traditional but popular policy adopted by several well-known cluster management systems, such as Yarn and Kubernetes. However, it usually performs poor due to its runtime-agnostic scheduling paradigm.
Picks the top-$G_j$ GPU with least execution time first. 
\subsubsection{Shortest Job First (SJF)}
an ideal policy to minimize the average JCT without preemption by prioritizing short-term jobs to overcome HOL blocking. It is impractical as it requires perfect job information which is impossible to attain.
\subsubsection{Tiresias \cite{tiresias2019}}
a preemptive policy that prioritizes least attained service jobs (i.e., consumed GPU numbers and training iterations). Under this policy, it helps short-term jobs escape from resource starvation and finish earlier without any prior information.
\subsubsection{Shortest Job First with First Fit Sharing (SJF-FFS)}
a sharing policy built upon SJF. It is similar to our proposed SJF-BSBF except that it does not search the best sharing configuration as SJF-BSBF but allocates the job to those GPUs that only have one job in a first fit manner if the free GPUs are not sufficient for the new job. This policy is a comparison baseline to validate the effectiveness of wisely sharing the GPUs in SJF-BSBF.
\subsubsection{Pollux \cite{aurick2021_pollux}}
the state-of-the-art elastic scheduler that adaptively adjust the GPU resources for each job to optimize the overall job performance as well as resource utilization. As explained in \cite{hu2023_lucid}, Pollux cannot guarantee no accuracy degradation for all models as it allows the scheduler to tune the training batch size, while our SJF-BSBF applies gradient accumulation to attain the same convergence as the original user specific batch size setting.

For physical experiments, we compare our SJF-BSBF with FIFO, SJF and Tiresias to demonstrate the advantages of resource sharing over those exclusive-mode policies. For simulation experiments, we also add Pollux, one of the state-of-the-art elasticity-based scheduler, to compare the sharing-based and the elasticity-based policies.

\textbf{Workload Settings:} We generate the workload similar to the Microsoft job trace \cite{jeon2019analysis}. More details about the Microsoft trace can be found in \cite{jeon2019analysis} and Appendix of \cite{tiresias2019}. For the physical experiments, considering that our testbed only has 4 nodes with 16 GPUs, we generate totally 30 DDL jobs by scaling down the original job trace. As job characteristics, such as the number of GPUs and the training iterations, we mostly follow the distributions of the real trace: 20 jobs using no more than 8 GPUs and 10 jobs using 12 or 16 GPUs. The training iteration of jobs varies from 100 to 5000. For the simulation experiments, we mainly follow the settings of Pollux \cite{aurick2021_pollux}. We randomly sample 240 jobs from the busiest period in the deep learning cluster traces published by Microsoft \cite{jeon2019analysis} and also annotate six DL tasks (BERT, CIFAR10, DeepSpeech2, ImageNet, NCF and YoloV3) used in Pollux to them. The settings of GPU numbers and training iterations also follow those of Pollux.

\subsection{Experimental Results on a Physical Cluster}
\begin{figure}[!ht]
	\centering
	\subfigure[JCT distributions of different scheduling policies in the physical experiments.]
	{
		\includegraphics[width=0.46\linewidth]{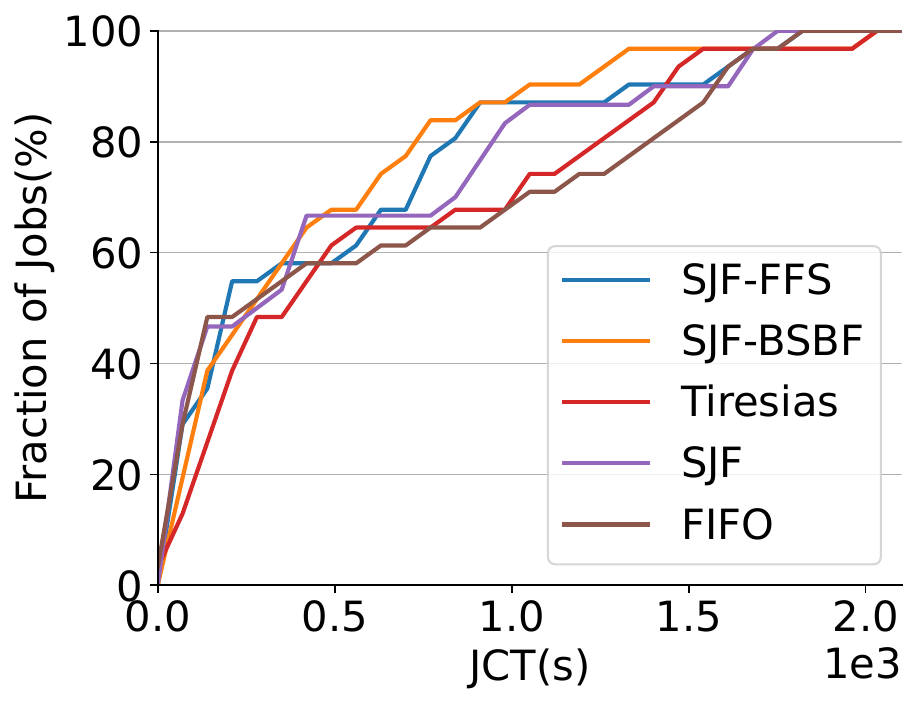}\label{fig:real_jct}
    }
	\subfigure[Average queuing time of different scheduling policies in the physical experiments.]
	{
        \includegraphics[width=0.46\linewidth]{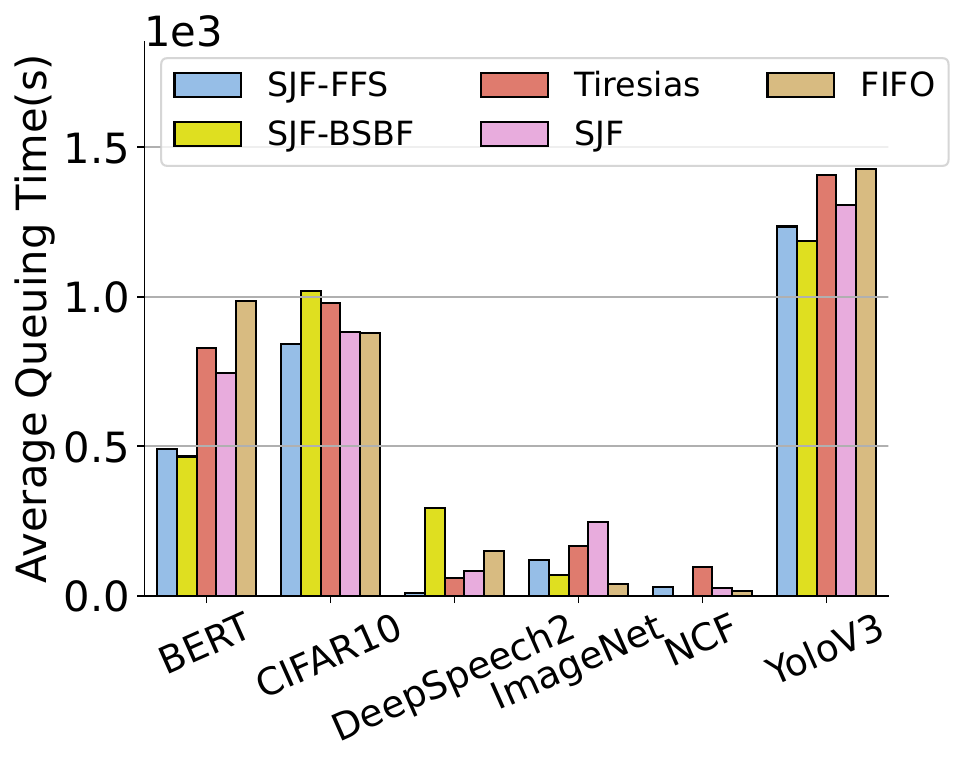}\label{fig:real_avg_queue}
        }
	\caption{Performance of different policies in physical experiments.}
	\label{fig:real_perf}
\end{figure}

\textbf{JCT Improvements:} 
Figure \ref{fig:real_jct} demonstrates the JCT distributions of the baseline workload using different scheduling policies. Nearly 80\% of jobs have no more than 0.75 hour of JCTs using our SJF-BSBF, while other algorithms only have less than 70\%. SJF-BSBF generally achieves the best performance. In Table \ref{tab:real_jct}, it is reported that SJF-FFS and SJF-BSBF, which allows the GPUs to be shared among jobs, have considerable performance improvements over other policies. In particular, SJF-BSBF achieves a 27\% lower average JCT than Tiresias. Besides, instead of allowing GPU sharing in a greedy manner, SJF-BSBF can adaptively select the best job combination as well as the scheduling point to avoid those job pairs that may bring down the overall performance, which outperforms SJF-FFS by 9\% in terms of JCT.

\textbf{Job Queuing Delay:} 
Figure \ref{fig:real_avg_queue} shows the average queuing time of different scheduling policies on different DDL job models. First, the queuing time of SJF-BSBF is generally lower than those heuristic policies with the exclusive GPU mode. For the model BERT, SJF-BSBF reduces the queuing time by nearly 44\% compared to Tiresias. Second, since SJF-FFS allows the jobs to share the GPUs in an aggressive manner, it generally has the lowest queuing time. However, as reported in Figure \ref{fig:real_jct} and Table \ref{tab:real_jct}, it usually leads to a longer JCT since some job pairs have a high interference ratio and subsequently hurt the overall performance. 

\begin{table}[!ht]
	\centering
	\caption{The makespan and average JCT run by physical experiments.}
	\label{tab:real_jct}
		\begin{tabular}{|c|c|c|} \hline
			Policy & Makespan (seconds) & Average JCT (seconds) \\ \hline\hline
			FIFO &  2317  &  662.6    \\ \hline
			SJF &  2319  &    609.55  \\ \hline
			Tiresias   &   2398   & 664.73  \\ \hline
			SJF-FFS    &  2176  & 530.36  \\ \hline
			SJF-BSBF  &  \textbf{2129}  & \textbf{483.16}  \\ \hline
		\end{tabular}
\end{table}

\subsection{Experimental Results on Large-Scale Simulations}
To verify the fidelity of our simulator, we also compare the results of physical experiments with simulations. We observe that the simulator can achieve the realistic experimental performance within 5\% relative percentage errors on both makespan and average JCT. This confirms the high fidelity of our simulator.

\begin{figure}[!ht]
	\centering
	\subfigure[JCT distributions of different scheduling policies in the simulation experiments.]
	{
		\includegraphics[width=0.46\linewidth]{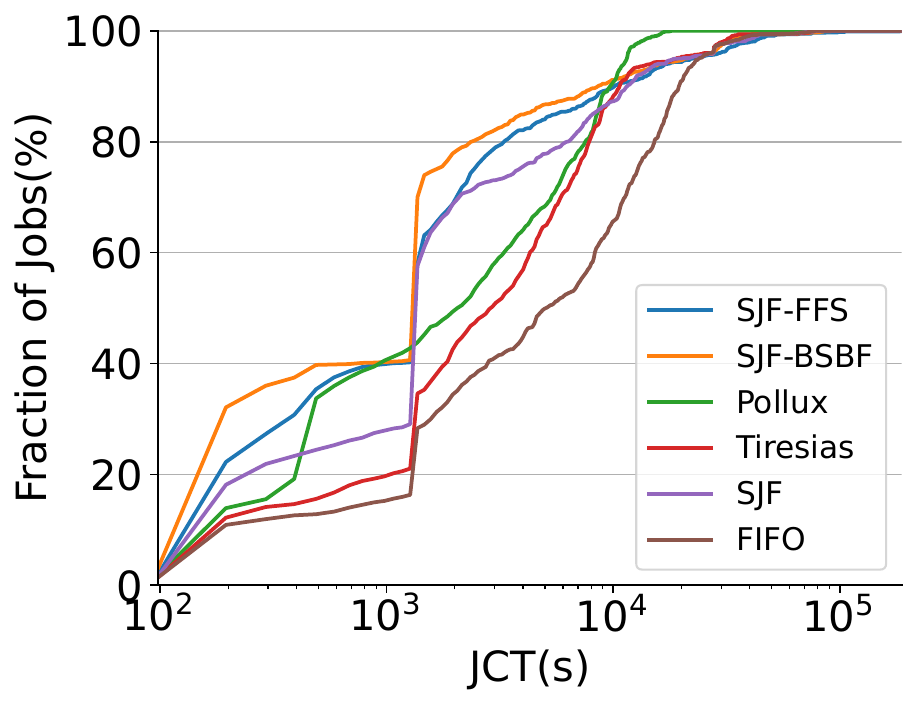}\label{fig:sim_jct}
    }
	\subfigure[Average queuing time of different scheduling policies in the simulation experiments.]
	{
        \includegraphics[width=0.46\linewidth]{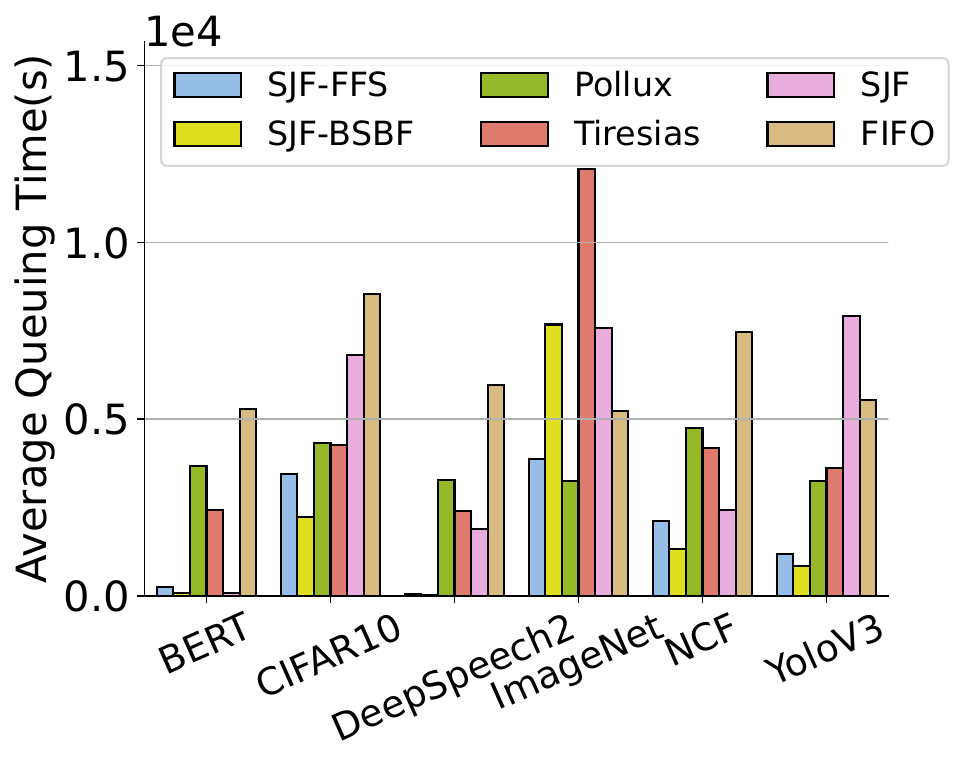}\label{fig:sim_avg_queue}
        }
	\caption{Performance of different policies in simulation experiments.}
	\label{fig:sim_perf}
\end{figure}

\textbf{JCT Improvements:} 
We first compare the JCTs of different scheduling policies on the standard simulation workload. In Figure \ref{fig:sim_jct}, it is evident that SJF-BSBF outperforms other policies. Nearly 40\% of jobs of SJF-BSBF achieves lower than 500 seconds of JCTs, reducing the average JCT of the shortest 40\% jobs by 37\% than Pollux. This demonstrates the preemption-free policy can even obtain better performance than the preemptive policy, such as Tiresias and Pollux.

Tables \ref{tab:sim_jct_1x} and \ref{tab:sim_jct_2x} present the performance of different scheduling policies for 240 jobs and 480 jobs, respectively. Jobs are characterized based on their requested number of GPUs, with those requiring more than 4 GPUs considered large, and others small.
For the workload of 240 jobs, SJF-BSBF demonstrates slightly better performance than the advanced policy Pollux. While large jobs under SJF-BSBF may experience longer JCTs than Pollux due to GPU sharing overhead, small jobs benefit significantly by potentially sharing GPUs with large jobs, resulting in markedly shorter queuing times compared to other preemption-free policies. This advantage is further accentuated as the number of jobs increases.
In Table \ref{tab:sim_jct_2x} with 480 jobs, SJF-BSBF enhances the average JCT by nearly 3 times compared to Pollux, primarily attributable to the reduction in queuing time for small jobs. Moreover, SJF-BSBF outperforms SJF-FFS by reducing the average JCT and queuing time by 17\% and 5.5\%, respectively.

\begin{table}[!ht]
	\centering
	\caption{Performance of large-scale and small-scale jobs by simulation. (240 jobs)}
	\label{tab:sim_jct_1x}
 		\addtolength{\tabcolsep}{-0.2pt}
		\resizebox{\linewidth}{!}{
		\begin{tabular}{ccccc} \hline
                Metrics (hrs) & Policy & All Jobs & Large Jobs & Small Jobs \\ \hline
			\multirow{5}{*}{Average JCT} & FIFO & 2.34 & 3.92 & 2.13 \\
                & SJF & 1.25 & 3.50 & 0.94 \\
                & Tiresias & 1.51 & 3.48 & 1.23 \\ 
                & Pollux & 1.04 & \textbf{1.96} & 0.91 \\ 
                & SJF-FFS & 1.23 & 4.29  & 0.81 \\
                & SJF-BSBF & \textbf{1.01} & 3.38 & \textbf{0.68} \\ 
                \hline
			\multirow{5}{*}{Average Queuing Time} & FIFO & 1.65 & 2.17  & 1.58 \\ 
                & SJF & 0.56 & 1.76 & 0.39  \\
                & Tiresias & 0.81 & 1.73 & 0.68 \\ 
                & Pollux & 1.00 & 1.21 & 0.97  \\ 
                 & SJF-FFS & 0.18 & 0.98 & 0.07 \\
                & SJF-BSBF & \textbf{0.14} & \textbf{0.89} & \textbf{0.03} \\ 
                \hline
		\end{tabular}
  }
\end{table}

\begin{table}[!ht]
	\centering
	\caption{Performance of large-scale and small-scale jobs by simulation. (480 jobs)}
	\label{tab:sim_jct_2x}
		\addtolength{\tabcolsep}{-0.2pt}
		\resizebox{\linewidth}{!}{
		\begin{tabular}{ccccc} \hline
                Metrics (hrs) & Policy & All Jobs & Large Jobs & Small Jobs \\ \hline
			\multirow{5}{*}{Average JCT} & FIFO & 7.52 & 10.06 & 7.08 \\
                & SJF & 2.55 & 7.00 & 1.90 \\
                & Tiresias & 5.09 & 8.25 & 4.59 \\
                & Pollux & 4.69 & \textbf{6.03} & 4.51 \\ 
                & SJF-FFS & 1.89 & 8.12  & 0.99 \\
                & SJF-BSBF & \textbf{1.57} & 6.13 & \textbf{0.88} \\ 
                \hline
			\multirow{5}{*}{Average Queuing Time} & FIFO & 0.73 & 4.06  & 0.21 \\ 
                & SJF & 1.87 & 5.35 & 1.35  \\
                & Tiresias & 4.41 & 6.60 & 4.04 \\ 
                & Pollux & 4.41 & 5.03 & 4.31  \\ 
                 & SJF-FFS & 0.73 & 4.06 & 0.21 \\
                & SJF-BSBF & \textbf{0.69} & \textbf{3.62} & \textbf{0.21} \\ 
                \hline
		\end{tabular}
  }
\end{table}

\textbf{Job Queuing Delay:}
Figure \ref{fig:sim_avg_queue} presents a comparison of the average queuing time among different scheduling policies for various DDL job tasks in simulation. Notably, the GPU sharing policies, namely SJF-BSBF and SJF-FFS, consistently yield lower queuing times compared to heuristic policies operating in exclusive GPU mode. Additionally, preemptive policies such as Tiresias and Pollux often exhibit longer queuing times attributable to job migration.

\begin{figure}[!ht]
	\centering
	\subfigure[Varying the workloads.]
	{
		\includegraphics[width=0.44\linewidth]{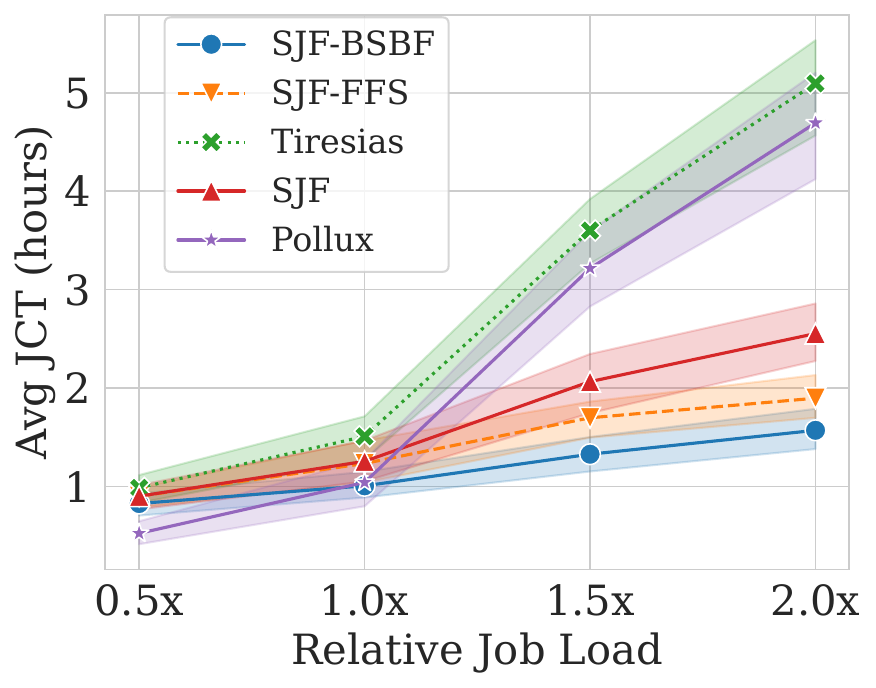}\label{fig:sim_abs_load}
    }
	\subfigure[Varying the interference ratios.]
	{
        \includegraphics[width=0.48\linewidth]{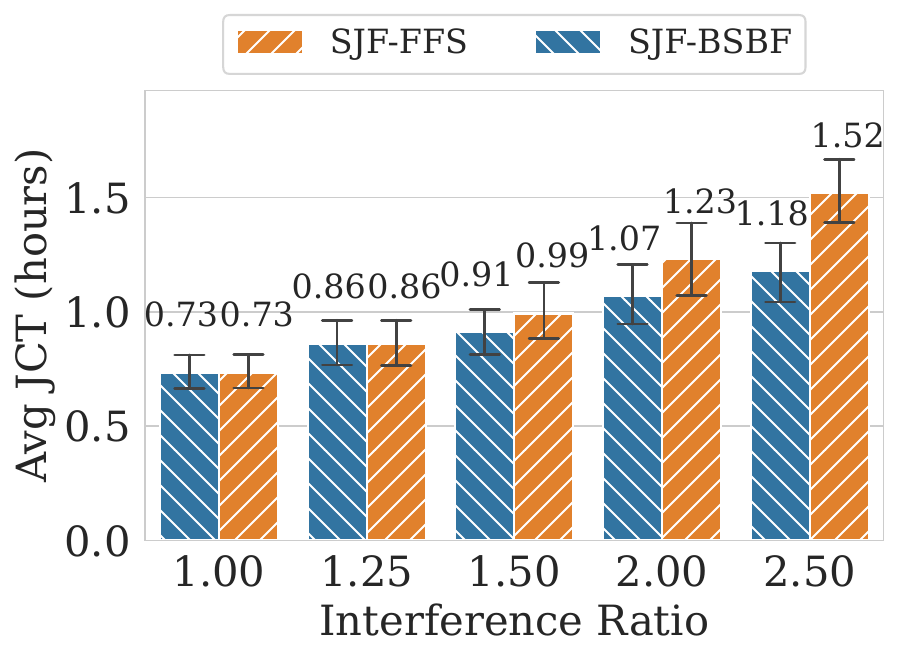}\label{fig:sim_abs_interfere}
        }
	\caption{Effects of various workloads and interference ratios.}
	\label{fig:sim_abs}
\end{figure}

\textbf{Sensitivity to job load:}
We compare the performance of our SJF-BSBF to other existing policies for increasing workload intensity in terms of job submission frequencies. We scale the baseline workload of 240 jobs by $0.5\times\sim2\times$, ranging from 120 jobs to 480 jobs. Figure \ref{fig:sim_abs_load} shows the results. An interesting phenomenon is that Pollux can have better performance than other policies when the job workload intensity is low. Pollux is more suitable for lighter workload intensity because its adaptive job batch size and resource scaling techniques are limited when clusters are overloaded, which meets the findings in \cite{hu2023_lucid, blox_2024}. However, when the workload increases, the GPU resources are rather insufficient so that Pollux cannot benefit from this strategy. Across all job workloads, our SJF-BSBF maintains relatively low improvements over other baseline policies since it allows the jobs to share the GPUs to shrink the job queuing time. 

\textbf{Impact of Different Interference Ratios:}
To evaluate the impact of different interference ratios on our GPU sharing policies, SJF-FFS and SJF-BSBF, we artificially inject various values for all the jobs sharing the same GPUs in the baseline simulation workload. Figure \ref{fig:sim_abs_interfere} shows the results. When the ratio is small ($\xi\le1.25$), which is the ideal scenario that sharing GPUs brings negligible overhead, SJF-BSBF tends to allow all the available sharing decisions as SJF-FFS, which results in the same performance. However, when the GPU sharing leads to severe slowdowns for the running jobs, our SJF-BSBF can get rid of those job pairs that may hurt the overall performance in SJF-FFS, which reduces the average JCT by 8\%$\sim$13\% when $\xi$ ranges from 1.5 to 2.0.

\section{Conclusion}\label{sec:conclusion}
In this paper, we delve into resource scheduling for DL jobs in a multi-tenant GPU cluster, where we harness GPU sharing capabilities to diminish job queuing time and enhance overall performance. 
We begin by formulating a DL scheduling model that accounts for GPU sharing among various jobs and employs gradient accumulation to surmount memory limitations while maintaining the job training accuracy. We then derive the optimal solution to schedule a job pair on the same set of GPUs and further design an efficient heuristic scheduling algorithm upon it to unleash the potential of GPU sharing in reducing the job queuing time and avoid serious interference with the running jobs. 
Extensive experiments, including physical implementations and simulations, were conducted to demonstrate the effectiveness of SJF-BSBF. Our findings reveal that the non-preemptive SJF-BSBF surpasses advanced preemptive policies like Tiresias and Pollux by leveraging GPU sharing techniques. Furthermore, identifying appropriate sharing settings is pivotal in mitigating severe degradation cases induced by high interference.

\section*{Acknowledgments}
This research was supported by the National Natural Science Foundation of China (No. 62302126, No. 62302123), the Shenzhen Science and Technology Program (No. RCBS20221008093125065, No. JSGGKQTD20221101115655027, No. JCYJ20220818102414030, No. KJZD20230923115113026, No. KJZD20230923114213027), and Guangdong Provincial Key Laboratory of Novel Security Intelligence Technologies (2022B1212010005). 

\bibliographystyle{IEEEtran}
\bibliography{main}

\end{document}